\documentclass[journal=preprint,manuscript=article]{achemso}

\usepackage[version=3]{mhchem} 
\usepackage{amsmath,amssymb}
\usepackage{graphicx}
\usepackage{float}
\usepackage{hyperref} 
\usepackage{xcolor}
\usepackage{siunitx}

\author{P.R. A de Oliveira}
\affiliation{Centro Brasileiro de Pesquisas Físicas, 22290-180, Rio de Janeiro, RJ, Brazil}
\email{hninofilho@gmail.com}
\author{P. Venezuela}
\affiliation{Instituto de Física, Universidade Federal Fluminense, Campus da Praia Vermelha,
Niterói, RJ, 24210-346, Brazil}
\author{F. Stavale}
\affiliation{Centro Brasileiro de Pesquisas Físicas, 22290-180, Rio de Janeiro, RJ, Brazil}
\author{J.A. Boscoboinik}
\affiliation{Center for Functional Nanomaterials, Brookhaven National Laboratory,
Upton, NY, 11973, USA}
\title{Insights into $CO_{2}$ activation on defective ZnS surfaces}


\begin{document}

\begin{abstract}
In this work, we investigate $CO_{2}$ activation on ZnS using Near Ambient-Pressure X-ray photoelectron spectroscopy measurements (NAP-XPS) and density functional theory calculations (DFT). Our NAP-XPS experiments reveal that $CO_{2}$ adsorbs onto a defective ZnS surface upon heating above $473 \ K$ in a $CO_{2}$ atmosphere (up to $0.55 \ mbar$). The $CO_{2}$ adsorption fingerprint is detectable even after cooling to room temperature under ultra-high vacuum.  Our DFT calculations suggest that $CO_{2}$ adsorption is energetically favorable on ZnS surfaces containing zinc vacancies, highlighting defect sites as key adsorption centers. Additionally, oxygen adsorption on a defective ZnS surface is exothermic, in contrast to the endothermic behavior observed on a defect-free surface. These findings contribute to a deeper understanding of defect-driven surface reactivity and may inform ZnS-based catalyst's design for $CO_{2}$ capture and reutilization. 
\end{abstract}

\section*{Introduction}

\ \ Over the past decade, global emissions of pollutant gases, such as C$O_2$, posing major environmental and energy-related challenges \cite{armaroli2007future}. Both academic and industrial effort has been devoted to exploring new routes of converting C$O_2$ into value-added products through catalytic processes such as $CO_{2}$ hydrogenation and electrochemical reduction reaction \cite{gao2023experimental, hu2021electrochemical}. In this regard, the development of high-efficient catalyst platforms for $CO_{2}$ capturing and reutilization technologies is essential \cite{collado2018unravelling}. 
Besides traditional oxide-based catalyst such as ceria oxide \cite{cheng2013carbon} and zinc oxide \cite{tang2013adsorption,chen2021controlling}, and most recently perovskite structures \cite{kim2024study}, transition metal chalcogenide have strong potential for reacting with $CO_{2}$ \cite{wang2021nanostructured}. In particular, zinc sulfide, due to its suitable electronic structure featuring a conduction band minimum that satisfies the reduction potential for $CO_{2}$ reduction reaction \cite{luo2023synthesis}. Although promissing due to its electronic configuration, the lack of active sites on pure ZnS systems hinders reactivity. 


The introduction of native defects is a promising strategy to increase the number of active sites \cite{hinuma2018density,zhong2025advances}. In particular, shallow-level defects such as cation vacancies can create localized electronic states that enhance the adsorption and activation of gas-phase molecules. These defect sites can facilitate the first step of a catalytic reaction by promoting adsorption. These adsorbed species may participate in downstream reactions, such as the synthesis of methanol, methane, or Formate from activated $CO_{2}$ species \cite{cai2023highly,prabhakar2024insights}. 

In addition to defect engineering, co-exposure strategies offer another route to enhance $CO_{2}$ adsorption and activation. For instance, CO has been shown to restructure catalyst surfaces by extracting surface atoms or altering the oxidation state of metal sites \cite{li2024oxygen2}. Similarly, co-exposure to O$_{2}$ may promote the removal of sulfur atoms from the ZnS surface, generating additional sulfur vacancies distinct from zinc vacancies. These newly created defect sites could further improve the ability of ZnS to adsorb and activate $CO_{2}$ \cite{gao2025sulfur,luo2023synthesis}. However, such strategies remain unexplored for zinc-blende (ZB) ZnS surfaces, particularly those featuring cationic (Zn) defects. Besides understanding whether $CO_{2}$ molecules could adsorbs on defective ZnS surfaces, fruitful insights of the activation mechanism could derive by in-situ monitoring  the intermediate species originated during the reaction between ZnS and $CO_{2}$ molecules. Yet, to our knowledge, this kind of investigation was not performed on ZnS surfaces until the present moment.

In this work, we investigate the mechanism of $CO_{2}$ adsorption on defective ZnS surfaces using a combination of near ambient pressure X-ray photoelectron spectroscopy (NAP-XPS) and density functional theory (DFT). Our study is divided into two main parts: Firstly, we analyze the thermodynamic conditions governing $CO_{2}$ adsorption by comparing the effects of temperature and $CO_{2}$ pressure. Secondly, we explore the role of mixed gas environments, specifically $CO_{2}$ + $CO$ and $CO_{2}$ + $O_{2}$, on adsorption behavior. Complementary DFT calculations are employed to provide atomic-scale insight into the adsorption mechanisms and to distinguish between the thermodynamical stability of C$O_{2}$ and $O_{2}$ adsorption  on pristine versus defective ZnS surfaces, confirming the cation vacancies provide active sites that turn the interaction energetically feasible. 

\section*{Experimental and Computational Methods}

\subsection*{Sample Preparation and Cleaning Procedure}

ZnS (001) oriented single crystals purchased from \textit{SurfaceNet GmbH} were cleaned using repeated cycles of Ar$^{+}$ sputtering and thermal annealing, following the same procedure described in our previous work \cite{de2025formation}. Briefly, the surface was sputtered using argon ions (\textit{E} = 600~eV, \textit{I}$_e$ = 10 mA) for 5 minutes, followed by annealing for 30 minutes at 1520 K. This protocol has been proven to reproducibly generate a Zn-deficient ZnS surface, as evidenced by the S 2p /Zn 3p relative  concentrations shown in the Supporting Information (Fig.S1). The cleaning of the sample was checked by UHV XPS measurements ensuring that  carbon, oxygen, nor other impurities were detected.  All elemental quantification were performed considering the respective inelastic mean free path (IMFP) of each core level. Prior to AP-XPS measurements, the surface was characterized under UHV conditions to ensure the absence of carbon or oxygen contaminants, thereby confirming an ideal surface for subsequent reactivity experiments.

\subsection{Near Ambient Pressure XPS Measurements}

Operando NAP-XPS experiments were carried out at a laboratory-based NAP-XPS system at \textit{Brookhaven National Laboratory (BNL)} featuring a  SPECS PHOIBOS NAP 150 hemispherical analyzer. The system comprises a multiple differential pumping stage between the main chamber (basal pressure $ < \ 3.10^{-10} \ mbar $) and the analyzer, allowing UHV conditions in the analyzer, even exposing the sample in the main chamber to a few mbar. In this system, the photoelectrons generated due to the incidence of a monochromatic Al-k$\alpha $ X-ray radiation (1486.6 eV) reach the electron energy analyzer after escaping through a small aperture of the order of $300 \ \mu m$. During NAP-XPS experiments, core-level spectra were acquired as fast as possible, employing a pass energy and step energy of 50 eV, keeping a rational balance between the signal-to-noise ratio while monitoring spectral evolution during the reaction. In all experiments, the sample was first exposed to the target gas mixture, then heated to the desired temperature (typically 573 K) and held for 60 minutes. Post the reactions,  UHV XPS spectra were acquired at room temperature. Survey and high-resolution spectra were collected with a pass energy of 50 eV and 30 eV, respectively.

\subsection{XPS Spectral Analysis}

All spectra were analyzed using the \textit{CasaXPS} software.  The Tougaard background was used with mixed Gaussian–Lorentzian (GL($x$)) functions, where $x = 40$ was found to minimize the residual error across all fits, to fitting the components from high-resolution spectra. All XPS spectra are shown without any binding energy
correction. Peak positions, full width at half maximum (FWHM), and constraints used to fit the data are reported in the Supporting Information (Table S1).

\subsection{DFT Calculations}

First-principles calculations were performed using the \textit{Quantum ESPRESSO} package \cite{giannozzi2009quantum}. The calculations were performed within the generalized gradient approximation (GGA), following the description of Perdew, Burke, and Enzenhorf \cite{perdew1996generalized}, employing ultra-soft pseudopotential \cite{vanderbilt1990soft} to solve the Kohn–Sham equations.  The kinetic energy of the wave functions and density charge cutoffs were 38 Ry and 456 Ry, respectively. The system was optimized by computing the equilibrium lattice parameter through calculations with a $6\times6\times3$ k-mesh for the Brillouin zone integration using a
Gaussian smearing function of $\sigma = 0.05 \ eV.$ The ZnS surface was modeled as a (2$\times$2) supercell of the (001) facet containing 80 atoms, separated by a vacuum spacing of 15 \AA{} to avoid fictitious periodic interactions. The bottom atomic layers were fixed to mimic bulk-like behavior, while the top two layers were allowed to relax fully during geometry optimization. A Zn-deficient model was created by removing one Zn atom from the topmost surface layer, resulting in a defect concentration below 5\%. All adsorption energies were computed using Grimme’s D2 van der Waals corrections \cite{grimme2010consistent}, which were essential for accurately describing interactions between the ZnS surface and adsorbates ($CO_{2}$, $O_{2}$). The adsorbates were optimized separately in a cubic box of 20 \AA{}. Structure optimizations were considered converged when the Helmman-Feyman forces on each atom were below 0.001/\AA{}. The adsorption energies of the adsorbates onto the ZnS surfaces were calculated as follows:

\begin{equation}
    E_{ads}(X) = E_{(ZnS+X)} - E_{ZnS} - E_{X} .
    \label{eqAds}
\end{equation}

In Eq.\ref{eqAds}, the terms $ E_{(ZnS+X)}$ and $E_{ZnS}$  denote the DFT total energy of the ZnS slab with and without the adsorbate $X$, respectively. The last term $ E_{X}$ is the total energy of the isolated $X$ adsorbate. All the structures are available in the Supporting Information.

\section*{Results and Discussion}
\subsection*{General features}

\begin{figure}[h]
    \centering
    \includegraphics[width=1.0\textwidth]{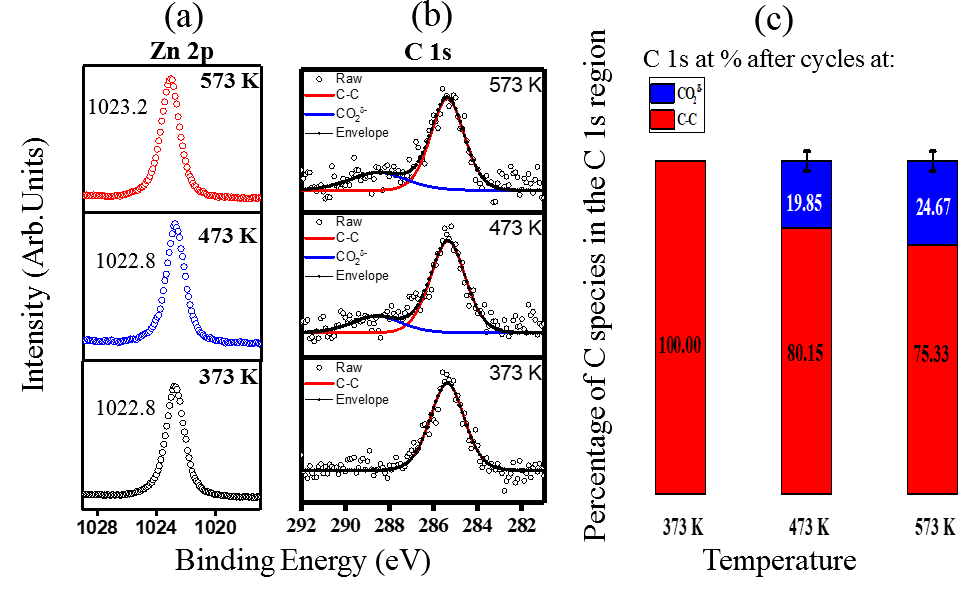}
    \caption{NAP-XPS operando analysis of (a) Zn 2p and (b) C 1s spectra as a function of the heating temperature; (c) Relative atomic concentration of the C 1s components as a function of the temperature. The error bar is indicated in black.
}
\label{paper4_fig1}
\end{figure}

The role of temperature in the adsorption of $CO_{2}$ molecules is assessed by analyzing the in situ reaction between $CO_{2}$ and the defective ZnS surface, as shown in Fig.\ref{paper4_fig1}. Fig. \ref{paper4_fig1}(a) displays the Zn 2p core-level spectra of the ZnS surface exposed to 0.55 mbar of $CO_{2}$ at different temperatures. Initially, at 373 K, the Zn 2p peak appears at the expected binding energy (B.E.) of 1022.8 eV. This binding energy remains nearly unchanged as the temperature rises to 473 K. However, at 573 K, a significant shift of 0.4 eV toward higher binding energy is observed. Such subtle modifications in the core-level position are typically associated with changes in the local chemical environment from which the photoelectrons are emitted. Under \textit{operando} conditions, a shift toward higher binding energy is generally interpreted as the onset of a chemical reaction, suggesting the interaction between $CO_{2}$ and the defective ZnS surface. Based on this behavior, we infer ZnS surface reactivity for $CO_{2}$ activation becomes thermodynamically favorable at 573 K. The O 1s evolution as a function of the temperature confirms that hypothesis. While at 473 K there is only a slight O 1s contribution likely deriving from the weak interaction of the adsorbate and the catalyst surface, the O 1s signal at 573 K reveals a new component at low binding energy, indicating the partial oxidation of the ZnS, as depicted in Fig.S2. 

Further insights into the $CO_{2}$ adsorption on ZnS are obtained by analyzing the C 1s core-level spectra under the same experimental conditions, as shown in Fig.\ref{paper4_fig1} (b). At 373 K, the C 1s spectrum is well described by a single component at 285.8 eV, attributed to C–C bonds in an $sp^{3}$ configuration \cite{patel2025initiation}. Notably, as the temperature increases, a satellite peak 3 eV away from the main component is observed.  This component is widely reported as a $CO_{2}$ activation fingerprint \cite{zhang2021exploring}, denoted as $CO_{2}^{\delta -}$ \cite{zhang2021exploring}. Interestingly, its intensity increases at 573 K, indicating that $CO_{2}$ adsorption could initiate at 473 K but becomes substantially more effective at 573 K. 

To gain quantitative insights, we analyzed the relative contributions of each C 1s component at each scenario discussed above, as shown in Fig.\ref{paper4_fig1}(c). At 373 K, no reaction is observed, and the C 1s signal is dominated by the C–C $sp^{3}$ component.  At higher temperatures, the $CO_2^{\delta -}$ component increases in intensity, particularly at 573 K, where it contributes to more than 24$\pm 3$ \% of the total C 1s area. This finding reinforces that 573 K is the optimal temperature for promoting $CO_{2}$ adsorption on the ZnS surface under the studied conditions.

\subsection*{On the activation mechanism}

The pathway for $CO_{2}$ activation on ZnS surface is not well established yet. As we discussed, there were no previous reports on ambient pressure experiments with ZnS-catalyst model. In that way, we based our analysis on similar candidates to gain insights into $CO_{2}$ activation mechanism. In most of NAP-XPS investigations, the high binding energy component is addressed to carbonates $CO_3$ features \cite{mehar2024ap,koitaya2016real}. In this context, it is assumed that $CO_{2}$ activation proceeds via a dissociative reaction that drives the formation of carbonates according to \cite{mehar2024ap}
\begin{equation}
    CO_{2_{(g)}} \xrightarrow{} CO_{g} + O_{chem} 
\end{equation}

\begin{equation}
    CO_{2_{(g)}} +O_{chem} \xrightarrow{} CO_{3_{(ads)}} 
\end{equation}

However, there are some tricky points to be disclosed. First, $CO_{2}$ weakly interacts with late transition metal atoms as zinc \cite{freund1996surface}. Secondly, this reaction presumes $CO_2$ molecules undergo full dissociation, which is not necessarily true when interacting with defective catalyst surfaces. Instead, a charge transfer between the molecule and some active sites derived from defects might take place, giving rise to an intermediate dissociation \cite{deng2008surface}.  In this context, we might speculate that oxygen from $CO_{2}$ interacts with zinc, which might lead to a partial oxidation of the ZnS surfaces, giving rise to the $CO_{2}^{\delta -}$ intermediate. Indeed, there are some species features that support our hypothesis. In particular, the shift of Zn 2p  components after evacuation and the shoulder at higher binding energy of C 1s component under \textit{operando} conditions. Regarding the first point, as we discussed in a previous work, the oxidation of ZnS surfaces leads to a core-level shift toward low binding energy after the reaction (i.e, after the system reaches its equilibrium). Notably, Zn 2p and S 2p are slightly shifted toward low binding energy after the $CO_{2}$ reaction (see Fig.S3 (a)-(b)). Moreover, we observed a third oxygen component after the reaction, assigned to oxygen bonding to hydroxyl species that form after the reaction. The $O_x$ species are shifted toward high binding energies, likely due to the interaction with water molecules derived post the reaction, as shown in (Fig.S3 (c)). The hydroxyl fingerprint is also detectable in the C 1s spectrum after evacuation, as will be discussed next. Regarding the second point, the intermediate $ CO_{2}^{\delta -}$ highlights the $CO_{2}$ activation mechanism. The reaction pathway is quite similar to the one described by Eq.(1), giving rise to a small oxygen contribution. The main difference is that the $ CO_{2}^{\delta -}$ intermediate decomposes onto ZnS surfaces, rather than $ CO_{2}$ molecules themselves, as occurs on Cu (111) surfaces \cite{deng2008surface,yang2010fundamental}. If we find carbonate species, we might speculate in a further step reaction as stated in Eq.(2), yielding carbonate species as well.  It is worth noting that although such interaction between adsorbate and substrate does occur, this charge transfer itself is not enough to promote the formation of a full zinc oxide layer. For this reason, we claim the $CO_{2}$ activation mechanism is a partial oxidation of ZnS, rather than a full $CO_{2}$ dissociation.  Interestingly, as we recently discussed, ZnS surface oxidation tends to increase as long as more oxygen species enter the ZnS lattice, giving rise to a thin ZnO layer. Notably, when the oxidation saturates, the ZnS's Zn LMM   drastically changes, becoming more ZnO-like \cite{de2025growth}. Nevertheless, it will not be possible to get a full ZnO/ZnS interface by oxidizing ZnS via $CO_{2}$, because the oxidative power of that molecule is much lower than $O_{2}$ \cite{mehar2024ap}. Indeed, the Zn LMM does not change after the $CO_{2}$ adsorption on ZnS (Fig.S3 (d))


\begin{figure}[h!]
    \centering
    \includegraphics[width=1.0\textwidth]{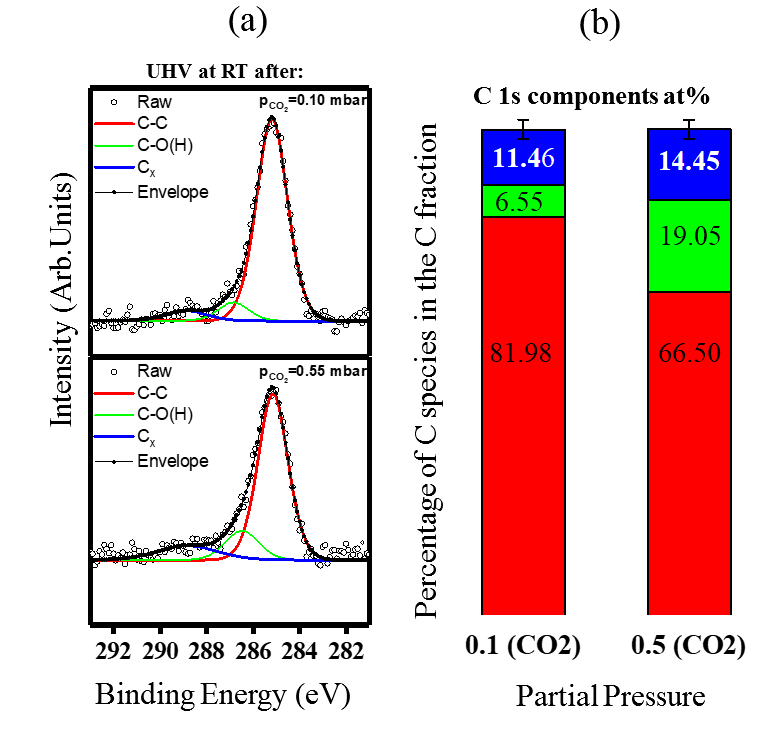}
    \caption{(a) UHV XPS measurements of C 1s components after NAP experiments at 0.10 mbar (top panel) and 0.55 mbar (bottom panel) of $CO_{2}$ ; (b) C 1s components relative concentration as a function of the pressure of $CO_{2}$ used in the experiments. In each NAP experiment, the sample was heated at 573 K. The relative concentration is with respect to the C 1s spectra collected under UHV conditions after the reaction. The error bar is indicated in black.
}
\label{pressure}
\end{figure}

Insights into the influence of $CO_{2}$ coverage on the activation mechanism were obtained by comparing the system's behavior under different $CO_{2}$ dosing conditions. As previously discussed, the surface was initially exposed to 0.55 mbar of $CO_{2}$. To evaluate the role of pressure, we then repeated the experiment under lower $CO_{2}$ exposure (0.1 mbar), maintaining the system at the identified optimal temperature of 573 K. Operando conditions XPS spectra do not reveal any subtle change in the C 1s components and contributions, as shown in Fig. S4. In particular, regarding the $CO_{2}^{\delta -}$ contribution: It represents 23.89 \% of the total area of the C 1s region  under 0.55 mbar of $CO_{2}$ pressure, and 22.80\% under 0.10 mbar of $CO_{2}$, which is  almost a negligible difference. On the other hand, UHV spectra after the reactions reveal some interesting features, as depicted in Fig.\ref{pressure}. Notably, there are still carbon species around 288 eV, in both scenarios. While operando conditions might give rise to carbonate or activated $CO_{2}^{\delta -}$ intermediate species, water molecules that inevitably desorb from the chamber and interact with the catalyst surface
after the reaction might recombine with these species, potentially forming Formate or related hydroxyl carbon species \cite{ma2023direct, zhou2014co2}. As we cannot precisely address the chemical nature of such a component, we will denote it as $C_{x}$ henceforth. The major difference between each scenario is the contribution of carbonyl (C-OH) species rather than $C_{x}$  species concentration, as noted in Fig.\ref{pressure} (b). While the green area represents only 6 $\pm 2$ \% of the overall C 1s signal after the experiment at low $CO_{2}$ pressure, its contribution increases to more than 19 $ \pm2$ \% after the experiments at higher $CO_{2}$ pressures. Conversely, $C_{x}$  species contribution increases by only circa 3 \% when comparing the C 1s spectra after reaction at $p_{CO_{2}} = 0.55$ mbar. These results suggest $CO_{2}$ coverage does not play a key role in the $CO_{2}$ activation on defective ZnS surfaces. Based on these findings, we selected 0.1 mbar of $CO_{2}$ at 573 K as the optimal target pressure and temperature conditions for $CO_{2}$ adsorption studies on ZnS surfaces, balancing effective activation with minimal impurity contributions. We now turn our attention to further mechanisms that allow the improvement of $CO_{2}$ activation on ZnS surfaces and, as a result, the oxidation of ZnS surfaces.

\subsection*{Mixing atmosphere: $CO_{2}$ + CO}

\begin{figure}[h!]
    \centering
    \includegraphics[width=1.0\textwidth]{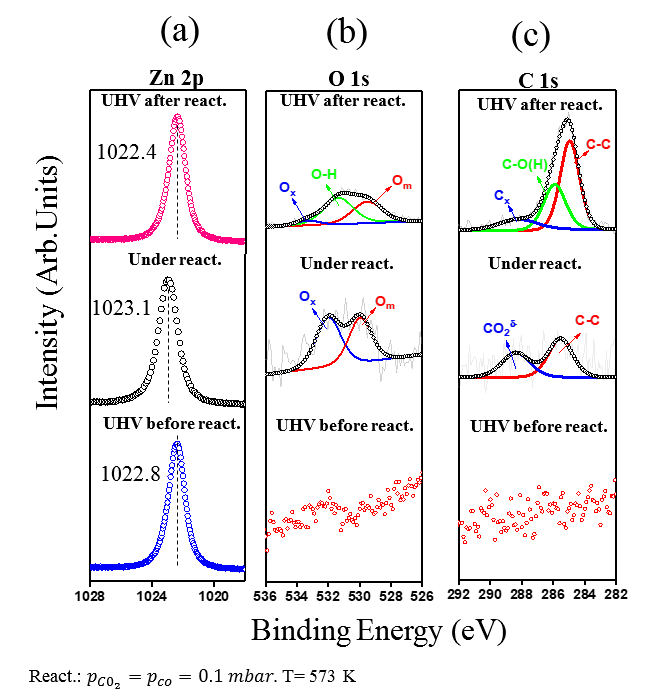}
    \caption{Zn 2p (a) C 1s (b) and (c) O 1s components before (bottom panel), under (middle panel) and after (top panel) NAP-XPS experiments with a mixed environment of $CO$ and $CO_{2}$ (0.1 mbar of each gas). The spectra before and after reaction were collected at room temperature under UHV conditions, while the spectra under reaction were acquired with the sample heated at 573 K.
}
\label{mixied-withCO}
\end{figure}

Inspired by past innovative works that overlooked the traditional metal-sensitive probe characteristic of CO, we explore its potential as an agent for restructuring catalyst surface \cite{zhou2022dynamical}. In this regard, Fig.\ref{mixied-withCO} displays the core level peaks of ZnS before, during, and after heating the system at 573 K in a combined CO and $CO_{2}$ atmosphere. The Zn $2p_{3/2}$ depicted in Fig.\ref{mixied-withCO} (a)  shifts as a function of the reaction status: Under UHV, before the reaction, it lies at 1022.8 eV, as expected in ZnS systems \cite{de2025formation,oliveira2024zinc}. During the reaction (Fig.\ref{mixied-withCO} (a) middle panel), it is shifted toward high binding energy. Ending, after the reaction ((Fig. \ref{mixied-withCO} (a) top panel), the component is slightly shifted toward low B.E, lying at 1022.6 eV. As discussed in the last section, changes in the core level B.E suggest changes in the chemical environment resulting from the adsorption of foreign species in the system. In our context, it is strictly related to $CO_{2}$ interacting with ZnS surfaces. The shift toward high binding energy at operando conditions highlights the reaction between adsorbates and the catalyst surface. The final shift after the reaction, on the other hand, reveals the final equilibrium configuration. In particular, this amount of shift is consistent with an initial oxidation of the ZnS surface \cite{de2025growth}. In this context, sulfate species are not expected, even after the reaction, due to their volatility. These features suggest the defective ZnS surface was activated by the mixed $CO_{2}$ and CO atmosphere. This hypothesis is well highlighted by analyzing the O 1s and C 1s peaks in the same scenarios mentioned before, as shown in Fig.\ref{mixied-withCO} (b) and (c), respectively. Before the reaction, there are no signals of carbon or oxygen species. This aspect changes significantly under operando conditions.  During the reaction (Fig.\ref{mixied-withCO} (b) middle panel), the O 1s component is well fitted by two oxygen species, highlighting the increase of the oxidative capability under a mixed $CO_{2}+CO$ atmosphere. The component at low binding energy, at 531 eV, is addressed to the interaction of oxygen with metal atoms, denoted as $O_m$. The high binding energy component at 532.8 eV, denoted by $O_{x}$, represents oxygen species in a dative-bonding configuration \cite{frankcombe2023interpretation}. This contribution is either related to the formation of $CO_3$ or the partial oxidation of ZnS through the $CO_{2}^{\delta -}$ activated species. At first glance, these values seem higher than the previous investigations regarding $CO_{2}$ fingerprint on catalyst surfaces. However, it is worth considering nothing but the relative position of the intermediate species rather than the absolute binding energies. Given the chemisorbed component in the O 1s peak, usually addressed to oxygen close to metal sites, activated $CO_{2}$ fingerprints arise 1.8 eV at higher binding energies, while carbonate features arise 2.2 eV offset \cite{deng2008surface}. In our case, the two oxygen components separation is 1.9 eV, indicating the $CO_{2}^{\delta -}$ intermediate nature of the peak. After the reaction (Fig.\ref{mixied-withCO} (b) top panel), a new component related to O-H group is obtained, due to the desorption of water molecules from the chamber. In this scenario, the $O_x$ component is slightly shifted toward high binding energy, likely due to the interaction with further hydrogen atoms from the $C_{x}$ species obtained after evacuation. More information on the reaction mechanism is obtained by analyzing the C 1s peak, as shown in Fig.\ref{mixied-withCO} (c). Under reaction conditions, it is well fitted by two components, one at 285 eV addressed to C-C bonds and a high-binding energy one at 288.6 eV attributed to the activated $CO_{2}$ intermediate species. These components are essentially the same as those obtained by exposing the catalyst to a $CO_{2}$ atmosphere alone, as discussed in the past section. The main difference in this mixed condition is the overall contribution of the activated species. While $CO_{2}^{\delta -}$ accounts for 24 $\pm 5$\% of the total C 1s area in a $CO_{2}$ atmosphere, it is responsible for 42.61 $\pm 3$\% in a mixed $CO_{2}$ + CO atmosphere (see Fig.S5 (a)). As long as the total pressure does not play a key role in the $CO_{2}$ activation, we might relate that improvement to the presence of further oxidative molecules, as CO. Although CO species are not expected to dissociate at ZnS surfaces, they might improve the activation of zinc atoms close to defect sites, partially changing their chemical state, introducing new active sites, or even promoting sub-surface zinc atoms to migrate to the top surface \cite{zhou2022dynamical,li2024oxygen2}, leading to an improved $CO_{2}$ activation. In this scenario, $CO_{2}$ molecules could also anchor close to oxygen species from CO, forming $CO_{3}^{-}$ intermediates, which are also located around the 289-289.5 eV region \cite{koitaya2016real,deng2008surface}. However, we are not able to disclose whether both species are contributing in this scenario. Following our fitting procedure, we deconvoluted the C 1s shoulder in a unique component at 288.7 eV, which is likely addressed to $CO_{2}^{\delta -}$. Alternatively, we could speculate that if only activated $CO_{2}^{\delta -}$ species were present, the contribution of remaining $C_x$ species obtained after evacuation would be higher, since there would be more activated species available for recombination. Although these species are indeed observed after the reaction, as shown in Fig.\ref{mixied-withCO} (c) (top panel), the contribution of this component is only 21.98 $\pm 3$\% of the total C 1s area. Therefore, we might speculate the higher activation component area observed during the reaction was likely driven by the formation of both activated $CO_{2}^{\delta -}$  and carbonate $CO_{3}^{-}$ intermediates. The last component is unstable and might desorb after the reaction, explaining why the final contribution after the reaction is as similar as post the reaction with $CO_{2}$ alone. 
\subsection*{Mixing atmosphere: $CO_{2}$ + $O_{2}$}

\begin{figure}[h!]
    \centering
    \includegraphics[width=1.0\textwidth]{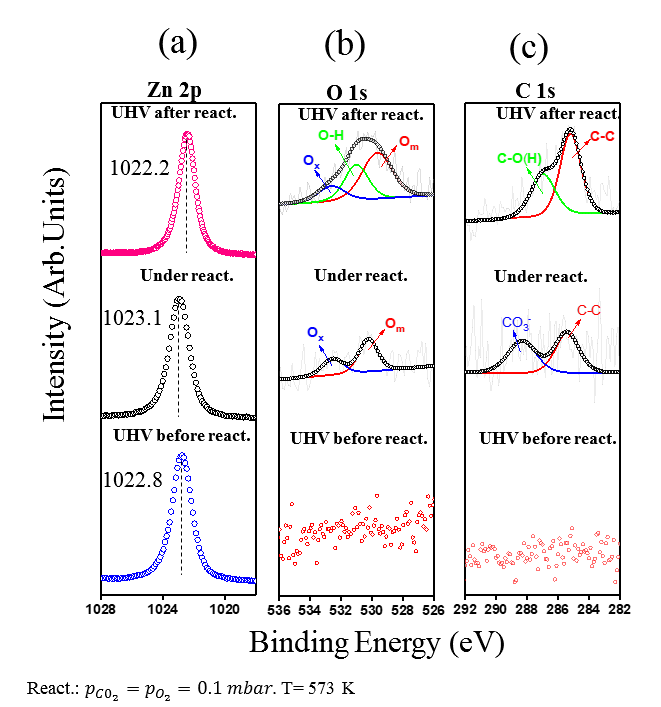}
    \caption{Zn 2p (a) C 1s (b) and (c) O 1s components before (bottom panel), under (middle panel), and after (top panel) NAP-XPS experiments with a mixed environment of $O_{2}$ and $CO_{2}$. The spectra before and after reaction were collected at room temperature under UHV conditions, while the spectra under reaction were acquired with the sample heated at 573 K.
}
\label{Mixed-CO2_O2}
\end{figure}

\newpage
To disclose the formation of activated $CO_{2}$, $CO_{3}$ or both species in the $CO_{2}$ adsorption on ZnS surface, we heated the system at 573 K in a $CO_{2} + O_{2}$ environment. Given that oxygen atoms display a higher oxidative capability than $CO_{2}$, the formation of a highly oxidized ZnS surface was expected, leaving $CO_{2}$ species to anchor on oxygen sites,  forming $CO_{3}^{-}$ intermediates.

Fig.\ref{Mixed-CO2_O2} (a) displays the Zn 2$p_{3/2}$ core level component before, during, and after exposing the system to a combined $CO_{2}$ and $O_{2}$ environment. Similar features regarding the surface reactivity were found: While during the reaction (Fig.\ref{Mixed-CO2_O2} (a) middle panel), the component is slightly shifted by 0.2 eV toward high binding energy with respect to the peak before the reaction (Fig.\ref{Mixed-CO2_O2} (a) top panel), it is shifted by 0.6 eV toward low binding energy after evacuation (top panel). This amount of shift after evacuation indicates the ZnS surface was highly oxidized, likely due to the presence of oxygen species. Indeed, due to both surface and electronic similarity of sulfur and oxygen species, oxygen atoms can be readily adsorbed on ZnS surface by occupying some surface sulfur sites. This straight oxidation via oxygen species seems more efficient than both $CO_{2}$ and $CO_{2}$ + CO atmosphere. The O 1s analysis displayed in Fig.\ref{Mixed-CO2_O2} (b) highlights this hypothesis. The sample before reaction does not feature any trace of oxygen species. On the other hand, under reaction, there are two notable oxygen components, at 531.3 eV and 533 eV, attributed to oxygen binding to zinc and dative-bonded oxygen species, respectively. These characteristics are very similar to those observed under the mixing of $CO_{2}$ and CO. However, the relative contribution of each species is quite different. Especially the $O_m /O_x$ ratio. While the $O_m$ contribution is 0.82 of the $O_x$ concentration when mixing $CO_{2}$ and CO, the $O_m /O_x$ ratio significantly increases to approximately 1.83 (see Fig.S5 (b)). This finding might be attributed to the full dissociation of $O_2$ species, which indeed display a higher oxidative capability than $CO_{2}$ molecules \cite{montemore2017o2}. After evacuation, besides the persistence of those components, an O-H contribution can be observed, as a result of the interaction of oxygen species with the desorbed water molecules after the reaction. As discussed, $O_x$ species are slightly shifted toward high binding energy in this scenario, likely due to the interaction with further hydrogen species, besides metal and carbon atoms. In this context, it is expected that the formation of a highly oxidized ZnS surface could lead to $CO_{2}$ species to anchor on oxygen sites in a co-adsorption mechanism rather than a straight interaction with metal species. This design is favorable for the formation of $CO_{3}^{-}$ species, as revealed in the C 1s peak analysis shown in Fig.\ref{Mixed-CO2_O2} (c). The flat line C 1s signal before the reaction is converted into an envelope that is deconvoluted into two main components under operando conditions: C-C species at 285.2 eV and a satellite peak at 289.3 eV. In the previous sections, we denoted the satellite component by $CO_{2}^{\delta -}$. However, since the interaction between oxygen and ZnS ($O_2 \rightarrow 2 \ O_{ads}$) proceeds faster than the interaction between $CO_2$ and ZnS, we might speculate that the interaction of $CO_{2}$ and the catalyst surface is mediated by oxygen species derived from the oxygen dissociation. Therefore, the satellite peak is most likely a $CO_{3}^{-}$-like component. Furthermore, the offset between the $sp^3$ carbon and the intermediate species is higher than 3 eV, which is the typical value for addressing the C 1s shoulder to $CO_{2}$ activated species. Also, even the absolute binding energy is in good agreement with previous investigations that reported the formation of $CO_{3}^{-}$ carbonate species. Indeed, the oxidation of ZnS via oxygen adsorption does not form a full ZnO layer \cite{de2025growth}, which could completely hinder the charge transfer from $CO_{2}$ to zinc atoms. However, the vast majority of the available active sites are likely from adsorbed oxygen species. Therefore, although the presence of further $CO_{2}^{\delta -}$ components cannot be ruled out, their contribution to that region, if it exists, would be almost negligible. This speculation is emphasized by the lack of $C_x$ species after evacuation, as depicted in the top panel of Fig.\ref{Mixed-CO2_O2} (c). As discussed, $CO_{3}^{-}$ species are unstable and likely desorb from the surface. If some activated $CO_{2}$ species were present, a minor though negligible $C_x$ feature would be detected after reaction. Therefore, we might speculate that $CO_{2} + O_{2}$ is favorable for $CO_{3}^{-}$ formation. 
\section{Theoretical Results}
\subsection*{Modeling the $CO_{2}$ interaction with ZnS surfaces}

\begin{figure}[h!]
    \centering
    \includegraphics[width=1.0\textwidth]{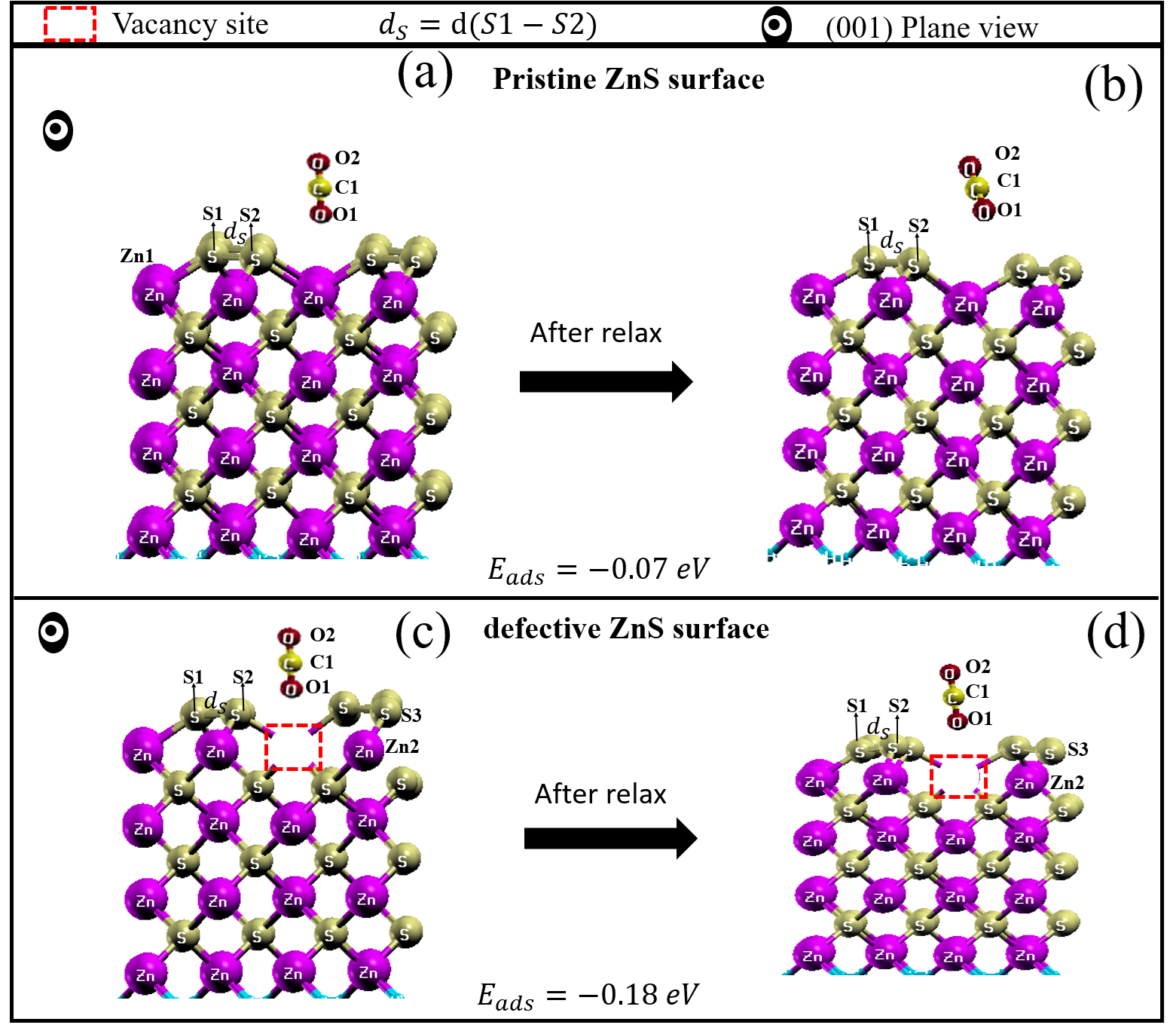}
    \caption{DFT modeling of ZnS pristine (a),(b), and doped (c),(d) before (left panel)  and after (right panel) interacting with $CO_{2}$. The unlabeled blue marks are hydrogen species used to passivate the bottom layers.
}
\label{DFT-CO2}
\end{figure}

\ \ To gain insights into the adsorption mechanism of $CO_{2}$ on ZnS surfaces, we investigated both the geometric and energetic aspects of $CO_{2}$ interaction with pristine and defective ZnS (001) surfaces, as illustrated in Fig.\ref{DFT-CO2}. Our models considered a sulfur-terminated ZnS (001) surface under zn-deficient conditions, in line with experimental preparation protocols. This polar surface is characterized by sulfur segregation at the outermost atomic layers, while the subsurface Zn atoms exhibit slight inward displacements upon structural relaxation. 

For the relaxed pristine structure, the vertical Zn–S bond distance at the topmost layer ($d(Zn1-S1)$) is \SI{2.45}{\angstrom}, and the lateral $d(S1–S2)$ distance between neighboring surface sulfur atoms is \SI{2.17}{\angstrom}. A $CO_{2}$ molecule, initially placed in its linear geometry (\SI{180}{\degree}), was positioned atop the surface with its molecular axis perpendicular to the slab, as depicted in Fig.\ref{DFT-CO2} (a). Upon optimization, no significant interaction between $CO_{2}$ and the pristine ZnS surface was observed, as noted in Fig.\ref{DFT-CO2} (b). Notably the distances between outer most zinc and sulfur species $d(Zn1-S1)$ as well as sulfur neighbors $d(S1–S2)$ were not disturbed, and the $CO_{2}$ molecule geometry remained nearly linear, slightly bending to \SI{179.5}{\degree}, suggesting minimal polarization. Moreover, the C=O bond length of \SI{1.17}{\angstrom} barely changed. The adsorption energy was calculated to be –0.07 eV, indicating the adsorption is slightly exothermic though with a extremely weak interaction between the adsorbate and the defect-free ZnS surface. The $CO_{2}$ binding energies were theoretically predicted to be about -0.1 eV on polar ZnO surfaces \cite{farias2013co2}. Bring this finding to the context of a polar ZnS structure, the calculated adsorption is below that binding energy, indicating $CO_{2}$ likely desorb rather than being activated on defect free ZnS surfaces, as experimentally suggested.

To explore the effect of native point defects on reactivity, we introduced a zinc vacancy ($V_{Zn}$) at the surface, which has been previously identified as the most thermodynamically stable defect under S-rich conditions \cite{de2025formation}. Upon removal of one Zn surface atom, as shown in Fig.\ref{DFT-CO2} (c)-(d), the surrounding sulfur atoms undergo localized relaxation. The distance between zinc and sulfur atoms in the vicinity of the vacancy ($d(Zn2-S3)$) slightly increases to \SI{2.47}{\angstrom}, reflecting an outward displacement and increased local sulfur density. Interestingly, upon exposure to $CO_{2}$, this bond length further extends to \SI{2.48}{\angstrom}, while the lateral $d(S1-S2)$ distance slightly contracts from \SI{2.14}{\angstrom} to \SI{2.13}{\angstrom}. These subtle though measurable relaxations imply a localized surface reconstruction in the presence of $CO_{2}$, likely induced by the altered electronic environment near to the vacancy site. The adsorption energy of $CO_{2}$ on the Zn-deficient surface was found to be –0.18 eV, noticeably more negative than on the pristine surface, and fitting the $CO_{2}$ binding energy aforementioned. This finding suggests the $CO_{2}$ adsorption on ZnS is thermodynamically favorable in a zinc-deficient scenario. The enhanced interaction suggests that zinc vacancies might promote partial activation of $CO_{2}$, possibly through electrostatic mild charge transfer, which gives rise to the $CO_{2}^{\delta -}$ activated intermediate, speculated through the experiments. Overall, this behavior is in good agreement to literature results showing that defect sites can serve as electron donors or centers of surface polarization, facilitating interactions with electrophilic molecules like $CO_{2}$ \cite{tang2013adsorption}. 

The  results derived from this simulation highlight that $CO_{2}$ interaction with pristine polar ZnS surface is very unlikely, while it might take place in the presence of native defects like zinc vacancies, via a weak van der Waals interaction, which agrees very well with the experimental results derived from NAP-XPS. This weak interaction confirms the hypothesis that $CO_{2}$ undergoes partial activation on ZnS rather than a full dissociation. Given the poor oxidative capability of $CO_{2}$ compared to oxygen molecules, the $CO_{2}$ activation might be mitigated when mixing both atmospheres. In this scenario, under operando conditions, $CO_{2}$ molecules likely interacts with oxygen atoms that might readily adsorbs on defective ZnS surface, as suggested by the experiments. This trend will be discussed next through modeling the oxygen adsorption on the same ZnS surfaces.
\subsubsection*{Modeling $O_{2}$ adsorption}

\begin{figure}[h!]
    \centering
    \includegraphics[width=1.0\textwidth]{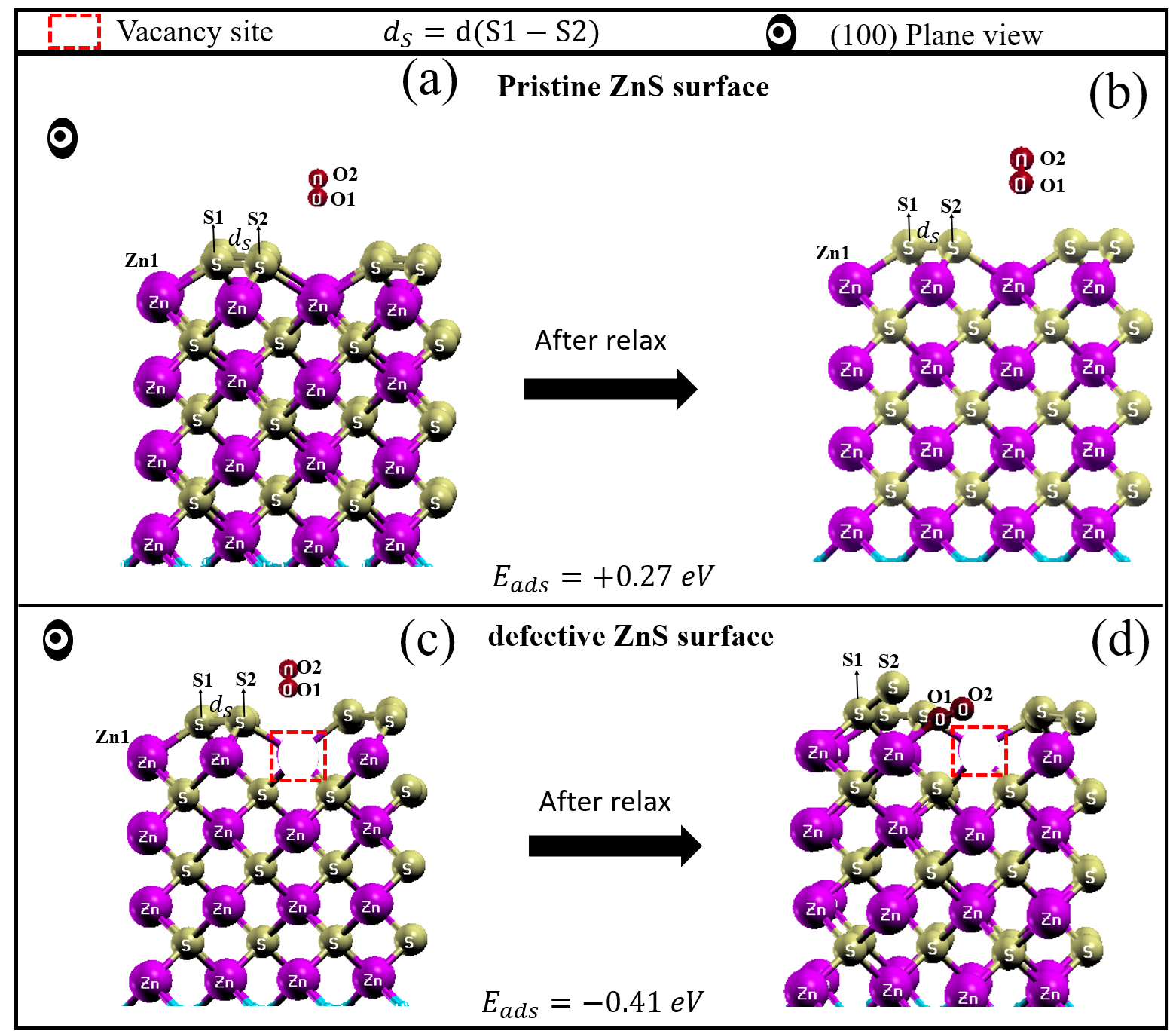}
    \caption{DFT modeling of ZnS pristine (a),(b), and doped (c),(d) before (left panel)  and after (right panel) interacting with $O_{2}$. The unlabeled blue marks are hydrogen species used to passivate the bottom layers.
}
\label{DFT-O2}
\end{figure}
We turn our attention for the interaction of oxygen and ZnS surfaces. The adsorption process was modeled by placing an oxygen molecule upward to the ZnS slab, positioned close to the outermost zinc atom.  The optimized oxygen distance $d(O1-O2)$ is \SI{1.23}{\angstrom}, in good agreement with previously results \cite{yang2018hybrid,han2022evaluation}. Fig.\ref{DFT-O2} (a)-(b) illustrates the structural features of oxygen adsorption on the defect-free ZnS (001) surface. As discussed in the past section, before reaction, the top layer sulfur atoms are separated by \SI{2.17}{\angstrom}, while the $d(Zn1-S1)$ distance is \SI{2.45}{\angstrom}. The main bond-length modification after relaxing the system was noted in the $d(O1-O2)$ separation, which was elongated to \SI{1.25}{\angstrom}. These findings would suggest that oxygen might react even with defect-free ZnS surfaces. However, one can note the quite high adsorption energy $E_{ads_{O_{2}}} = + 0.27 eV $. Therefore, oxygen adsorption on defect-free ZnS is in fact an endothermic process, which is not thermodynamically favorable.

The scenario is dramatically changed in the presence of a zinc vacancy, as revealed by Fig.\ref{DFT-O2}(c)-(d). Although $d(Zn1-S1)$ remains similar to that of the pristine surface, the lateral sulfur distances $d(S1-S2)$ decrease by \SI{0.07}{\angstrom} . This contraction is indicative of a localized sulfur segregation driven by the relaxation of surface atoms near the vacancy site. In addition, the $d(O1-O2)$ increases to \SI{1.32}{\angstrom}. This increase is often addressed to a dissociative chemisorption mechanism \cite{van2024o2}. As noted in Fig. \ref{DFT-O2} (d), the oxygen molecule indeed sticks on the ZnS surface with a lower adsorption energy - $E_{ads} = -0.41 \ eV $. This fact explains why oxygen atoms are indeed readily adsorbed on defective ZnS, as we observed experimentally. The mechanism behind such a chemisorption is explained as follows: 1) Sulfur atoms close to the vacancy sites move upward to the ZnS surface, breaking or weakening their bonds with neighboring Zn atoms. This structural rearrangement leads to the reduction of $d(S1-S2)$; 2) Next, the oxygen molecule is tilted, shifting laterally to interact with an uncoordinated Zn atom near to vacancy sites. This helps for compensating the dangling bond created by sulfur displacement, leading to a polarized $O_{2}$ configuration with an extended O-O distance. At this energetically favorable configuration, the bond strength is likely reduced compared to molecular $O_{2}$ species, which suggests oxygen might undergo dissociation ($O_2 \rightarrow 2 O_{ads}$) depending on the particular environment that they will be subjected to. Once all DFT calculation were carried at 0 K and without the presence of further adsorbate, we were not able to visualize the full dissociation of the chemisorbed oxygen species, which likely takes place by heating the system at 573 K as experimentally suggested.
\section*{Conclusion}

In this work we investigated the $CO_{2}$ adsorption on defective ZnS surfaces. By combining AP-XPS and DFT calculations, our results offer the first detailed insight into the surface chemistry of $CO_{2}$ on defective ZnS under near-operando conditions. Firstly, by heating the sample in a $CO_{2}$ environment, we determined 573 K as the ideal temperature for the $CO_{2}$ adsorption.  Secondly, working with two different pressures, we did not observe huge modifications on the $CO_{2}$ fingerprints, which suggests the $CO_{2}$ coverage does not play a role in the $CO_{2}$ adsorption efficiency.  In this scenario, we noted a shift on the core-level peaks, the rise of two oxygen components, and the remarkable remaining of the $CO_{2}$ fingerprint that suggests $CO_{2}$ partially oxidized ZnS surface, giving rise to final yields that seem stable even after evacuation. Next, mixing $CO_{2}$ with CO and $O_{2}$, we speculate on the nature of the satellite peak usually attributed to carbonate $CO_{3}^{-}$ intermediate. While the mixing with CO allow for a co-existence of both activated $CO_{2}$ (likely derived by the interaction between $CO_{2}$ and zinc) and $CO_{3}^{-}$-like species (likely derived by the interaction between $CO_{2}$ and oxygen from CO), the mixing with oxygen hinder the $CO_{2}$ activated component. This difference is attributed to the higher oxidative capability of O$_{2}$. Neither CO nor $CO_{2}$ molecules are expected to dissociate onto the ZnS surface fully. On the other hand, oxygen species readily chemisorb on ZnS surfaces, giving rise to an oxidative layer. In this scenario, all $CO_{2}$ molecules that could interact with ZnS, would make it only via a co-adsorption mechanism, mediated by these adsorbed oxygen species, which gives rise to $CO^{3-}$- like intermediate species. These species are not stable and leave the ZnS surface.  Our DFT calculations confirm $CO_{2}$ weakly interacts with ZnS surface, likely mediated by Van der Waals interactions. Conversely, oxygen species chemisorb on defective ZnS surfaces, likely dissociating, as suggested by the huge increase in the O-O bond length. Further investigations exploring the Gibbs free energy of the intermediate species might improve the knowledge of the activation mechanism in future work. Also, we might speculate if the co-adsorption of $CO_{2}$ close to oxygen sites give rise to unstable $CO_{3}^{-}$-like intermediates, CO molecules might give rise to $CO_{2}^{\delta -}$-like components that seem more stable, interacting with water molecules and remaining stable even after evacuation. Overall, our findings provide the first insights into the adsorption mechanism of $CO_{2}$ on defective ZnS and suggest this system as a promising platform for exploring the formation of methanol by combining the remaining products with hydrogen species.

\begin{acknowledgement}
The authors thank the Conselho Nacional de Desenvolvimento Científico e Tecnológico (CNPq), and the
Fundaçao de Amparo a Pesquisa do Estado do Rio de Janeiro
(FAPERJ) for financial support. P.R.A.d.O. and P.V. also
acknowledge the Centro Nacional de Processamento de Alto
Desempenho (CENAPAD-SP) for providing computational
resources.
\end{acknowledgement}

\bibliography{References-paper}

\end{document}